\documentclass[aps,a4paper,superscriptaddress,twocolumn]{revtex4}
\usepackage{tensor}
\usepackage{graphicx}
\usepackage{amsmath}
\usepackage{amssymb}
\usepackage{enumerate}
\usepackage{subfigure}
\usepackage{tabularx}
\usepackage[colorlinks=true, pdfstartview=FitV, linkcolor=blue, citecolor=red, urlcolor=black, breaklinks=true]{hyperref}
\newcommand{\be}{\begin{equation}}
\newcommand{\ee}{\end{equation}}
\newcommand{\ben}{\begin{eqnarray}}
\newcommand{\een}{\end{eqnarray}}
\newcommand{\bes}{\begin{subequations}}
\newcommand{\ees}{\end{subequations}}
\def\bal#1\eal{\begin{align}#1\end{align}}

\newcommand{\LL}{{\mathcal L}}

\newcommand{\veps}{\varepsilon}
\begin{document}
\title{Small and hollow magnetic monopoles}

\author{D. Bazeia}
\affiliation{Departamento de F\'\i sica, Universidade Federal da Para\'\i ba, 58051-970 Jo\~ao Pessoa, PB, Brazil}

\author{M.A. Marques}
\affiliation{Departamento de F\'\i sica, Universidade Federal da Para\'\i ba, 58051-970 Jo\~ao Pessoa, PB, Brazil}

\author{Gonzalo J. Olmo}
\affiliation{Departamento de F\'isica Te\'orica and IFIC, Centro Mixto Universidad de Valencia - CSIC. Universidad de Valencia, Burjassot-46100, Valencia, Spain}
\affiliation{Departamento de F\'\i sica, Universidade Federal da Para\'\i ba, 58051-970 Jo\~ao Pessoa, PB, Brazil}

\begin{abstract}
We deal with the presence of magnetic monopoles in a non Abelian model that generalizes the standard 't~Hooft-Polyakov model in three spatial dimensions. We investigate the energy density of the static and spherically symmetric solutions to find first order differential equations that solve the equations of motion. The system is further studied and two distinct classes of solutions are obtained, one that can also be described by analytical solutions which is called small monopole, since it is significantly smaller than the standard 't~Hooft-Polyakov monopole. The other type of structure is the hollow monopole, since the energy density is endowed with a hole at its core. The hollow monopole can be smaller or larger than the standard monopole, depending on the value of the parameter that controls the magnetic permeability of the model. 
\end{abstract}

\maketitle

%%%%%%%%%%%%%%%%%%%%%%%%%%%%%%%
\section{Introduction}

In classical field theory, topological structures usually emerge as static solutions of the equations of motion that describe the system. Among the most known topological structures that appear in field theory are the kinks, vortices and magnetic monopoles. They are one-, two- and three-dimensional objects, respectively, and find several applications in high energy physics, to describe phase transitions and other features, and in condensed matter, where they may be used to describe specific properties of superconductors and magnetic materials; see, e.g., Refs.~\cite{vilenkin,manton,S,weinberg,fradkin}.

Due to their topological nature, these structures are usually stable under small fluctuations. In general, kinks only require a single real scalar field and the global $Z_2$ symmetry to be generated. To study vortices, one has to consider scalar field and an Abelian gauge field that are minimally coupled and evolve controlled by the local $U(1)$ symmetry. Magnetic monopoles are much more intricate and, to deal with them, one may consider a triplet of scalar fields coupled to a non Abelian gauge field through a local $SU(2)$ symmetry; see, e.g., Refs.~\cite{thooft,polyakov}. 

The inclusion of extra degrees of freedom and the corresponding enlargement of symmetries have allowed for the presence of new features for vortices and monopoles. In the recent years, in Ref.~\cite{S1}, for instance, the author showed that conventional vortices can acquire non Abelian moduli localized on their world sheets, and in Ref.~\cite{S2} the system may be engineered to appear with a cholesteric vacuum in order to admit topologically stable vortices with additional features. More recently yet, in Ref.~\cite{vortexint} we have studied vortices in a model where the $U(1)$ symmetry is enlarged to become $U(1)\times Z_2$, with the magnetic permeability modified and controlled by an additional neutral scalar field which acts as a source field. This modification gave rise to interesting solutions of vortices with internal structure. A similar feature has recently appeared with monopoles in a $SU(2)\times Z_2$ model \cite{monopoleint}, where the additional neutral scalar field drives not only the magnetic permeability, but also the covariant derivative contribution due to the triplet of scalar fields, giving rise to monopoles with nontrivial internal structure.

In this work we focus on magnetic monopoles, extending the standard model originally studied in Refs.~\cite{thooft,polyakov} to the novel context that we explain in the next section, watching with special attention for the fact that monopoles are also important in condensed matter, in particular in magnetic materials generically known as spin ice \cite{SI1,SI2,SI3}, capable of supporting exotic magnetic structures that may fractionalize into monopoles. Another motivation comes from the possibility that magnetic monopoles in specific materials \cite{SI4} can support an electric dipole and, in this sense, may be endowed with internal structure. We also think that the current investigation is of interest to describe other possibilities, such as the enlarged system that couples a superconductor to a superfluid recently investigated in \cite{two} and also, the presence of color-magnetic defects \cite{dense} recently found in dense quark matter, that are energetically preferred in the parameter regime relevant for compact stars.

The presence of magnetic monopole with internal structure has motivated us to search for simpler models, capable of supporting localized objects with internal structure similar to monopoles. We concentrate on the issue without modifying the symmetry of the standard model, so we study generalized models similar to the ones proposed in Refs.~\cite{mono1,mono2}, with the standard $SU(2)$ symmetry, as in the pioneer works \cite{thooft,polyakov}. The investigation follows the lines of Ref.~\cite{monopoleint}, and it is also motivated by \cite{kvm}, where several generalized models that support kinks, vortices and magnetic monopoles were studied. An interesting feature of the study is the presence of first order differential equations that solve the equations of motion, as in the Bogomol'nyi-Prasad-Sommerfield (BPS) procedure firstly presented in \cite{ps,bogo}.    

To comply with this, we organize the work as follows: In Sec.~\ref{sec2} we introduce the generalized model and investigate the presence of first order differential equations that solve the equations of motion that describe the system. Based on this important simplification, in Sec.~\ref{sec3} we present some distinct models for magnetic monopoles, one that admits new analytical solutions that engender a non standard profile, which we call the small monopole, and others, presenting solutions with energy density with a hole in its core, which we call hollow monopoles. Finally, in Sec.~\ref{sec4} we present our conclusions and perspectives. 

%%%%%%%%%%%%%%%%%%%%%%%%%%%%%%%%%%%%
\section{Procedure}\label{sec2}

We work in $(3,1)$ flat spacetime dimensions with the Lagrangian density
\be\label{lmodel}
\LL = - \frac{P(|\phi|)}{4}F^{a}_{\mu\nu}F^{a\mu\nu} -\frac{M(|\phi|)}{2} D_\mu \phi^a D^\mu \phi^a - V(|\phi|).
\ee
In the above equation, $\phi^a$ is a triplet of real scalar fields coupled to the vector field $A^a_\mu$ under a $SU(2)$ symmetry. We also have the covariant derivative with respect to the scalar fields denoted as $D_\mu \phi^a = \partial_\mu\phi^a + g\,\veps^{abc}A^b_\mu\phi^c$ and the electromagnetic field strength tensor as $F^a_{\mu\nu} = \partial_\mu A^a_\nu - \partial_\nu A^a_\mu + g\,\veps^{abc}A^b_\mu A^c_\nu$.  Also, $g$ stands for the coupling constant, the indices $a,b,c=1,2,3$ are related to the $SU(2)$ symmetry of the fields and the greek letters $\mu,\nu=0,1,2,3$ denote the spacetime indices. We use the metric tensor $\eta_{\mu\nu} = \textrm{diag}(-,+,+,+)$ and natural units, $\hbar=c=1$. 

In this model, $P(|\phi|)$ is a function that generalizes the magnetic permeability and $M(|\phi|)$ modifies the dynamics of the scalar field. In this work we consider $P(|\phi|)$ and $M(|\phi|)$ non negative functions that depend only on the scalar fields, with $|\phi|^2=\phi^a\phi^a$. The main motivation is to allow the triplet of real scalar fields to modify the medium where the vector and scalar fields stand. This is the generalized model we want to study, keeping an eye open for the possibility to provide a continuum field theory description for the exotic magnetic materials collectively called spin ice, which are known to support localized structures similar to magnetic monopoles \cite{SI1,SI2,SI3,SI4}.  

One may vary the action associated to the Lagrangian density in Eq.~\eqref{lmodel} with respect to the scalar and vector fields to get
\bes\label{geom}
\begin{align}
 D_\mu\left(M D^\mu \phi^a\right) &=\frac{P_{\phi^a}}{4}F^{b}_{\mu\nu}F^{b\mu\nu} + \frac{M_{\phi^a}}{2} D_\mu \phi^b D^\mu \phi^b +  V_{\phi^a},\\ \label{meqsc}
 D_\mu\left(PF^{a\mu\nu}\right) &=  gM\,\veps^{abc}\phi^b D^\nu \phi^c,
\end{align}
\ees
where $D_\mu F^{a\mu\nu} = \partial_\mu F^{a\mu\nu} + g\,\veps^{abc}A^b_\mu F^{c\mu\nu}$ and $V_{\phi^a} = \partial V/\partial\phi^a$.
In order to seek for monopoles, we take static configurations with $A_0=0$ and suppose that
\be\label{ansatz}
\phi^a = \frac{x_a}{r} H(r) \quad \text{and} \quad A_i^a = \veps_{aib}\frac{x_b}{gr^2}(1-K(r)),
\ee
with the boundary conditions
\be\label{bcond}
\begin{aligned}
H(0)&=0, & K(0)&=1, \\
\lim_{r\to\infty}{H(r)} &= \pm\eta, & \lim_{r\to\infty}{K(r)} &= 0.
\end{aligned}
\ee
In this case, the equations of motion \eqref{geom} become
\bes\label{geomansatz}
\begin{align}
\frac{1}{r^2}\left(r^2M H^\prime\right)^\prime &= \frac{2M H K^2}{r^2} +\frac{P_H}{2} \left(\frac{2{K^\prime}^2}{g^2r^2} + \frac{(1-K^2)^2}{g^2r^4}\right)\nonumber\\
&\hspace{4mm} +\frac{M_H}{2}\left({H^\prime}^2 + \frac{2H^2K^2}{r^2}\right)  + V_{H},   \\
r^2\left(P K^\prime\right)^\prime &= K\left(M g^2r^2H^2 -P\,(1-K^2)\right),
\end{align}
\ees
where the prime denotes the derivative with respect to $r$. The above equations are of second order and present couplings between the functions $H(r)$ and $K(r)$. 

In general, the equations of motion are very hard to be solved, so we seek for first order equations that are compatible with them. To do this, we use the BPS procedure to minimize the energy \cite{ps,bogo} of the solutions. The energy density can be calculated standardly; we use \eqref{ansatz} to get
\be\label{rho}
\begin{aligned}
\rho &= \frac{P(|H|)}{2} \left(\frac{2{K^\prime}^2}{g^2r^2} + \frac{(1-K^2)^2}{g^2r^4}\right)\\
     &\hspace{4mm} + \frac{M(|H|)}{2}\left({H^\prime}^2 + \frac{2H^2K^2}{r^2}\right) + V(|H|).
\end{aligned}
\ee
Although it is hard to implement the BPS procedure for arbitrary $P$ and $M$, we have been able to achieve this goal under the choice $M(|H|)=1/P(|H|)$. This is a limitation, but we still have room to choose the generalized permeability adequately. In this case, the energy density becomes
\be
\begin{aligned}
\rho &=\frac{P(|H|)}{2} \left(\frac{H^\prime}{P(|H|)} \mp \frac{1-K^2}{gr^2}\right)^2 \\
     &\hspace{4mm} + P(|H|)\left( \frac{K^\prime}{gr} \pm \frac{HK}{rP(|H|)}\right)^2 + V(|H|) \\
     &\hspace{4mm} \pm \frac{1}{r^2}\left(\frac{\left(1-K^2\right)H}{g}\right)^\prime.
\end{aligned}
\ee
Now, we can discard the potential to write
\be
\begin{aligned}
\rho &= \frac{P(|H|)}{2} \left(\frac{H^\prime}{P(|H|)} \mp \frac{1-K^2}{gr^2}\right)^2 \\
     &\hspace{4mm} + P(|H|)\left(\frac{K^\prime}{gr} \pm \frac{HK}{rP(|H|)}\right)^2 \pm \frac{1}{r^2}\left(\frac{\left(1-K^2\right)H}{g}\right)^\prime.
\end{aligned}
\ee
It is straightforward to see that the two first terms are non-negative. Therefore, the energy is bounded, i.e., $E\geq E_B$, where
\be\label{ebogo}
E_B = \frac{4\pi\eta}{g},
\ee
is known as the Bogomol'nyi bound. If the solutions satisfy the first order equations
\bes\label{fom}
\bal
H^\prime &=\pm \frac{P(|H|)(1-K^2)}{gr^2},\\
K^\prime &=\mp \frac{gHK}{P(|H|)},
\eal
\ees
the energy is minimized and the Bogomol'nyi bound is attained; $E=E_B$, as given by Eq.~\eqref{ebogo}. One can show that the first order equations \eqref{fom} are compatible with the equations of motion \eqref{geomansatz} for $V(|\phi|)=0$. The pair of equations with the upper signs are related to the lower signs ones through the change $H(r)\to -H(r)$.

The presence of the above first order equations allows us to write the energy density in Eq.~\eqref{rho} as
\be\label{rhom}
\begin{split}
	\rho &= \frac{2P(|H|){K^\prime}^2}{g^2r^2} + \frac{{H^\prime}^2}{P(|H|)}\\
	       &= \frac{2H^2K^2}{P(|H|)\,r^2} + \frac{P(|H|)(1-K^2)^2}{g^2r^4}
\end{split}
\ee
As it was shown by the Bogomol'nyi procedure, we see from Eq.~\eqref{ebogo} that the energy is fixed, independently of the function $P(|H|)$ one chooses to define the model. 

%%%%%%%%%%%%%%%%%%%%%%%
\begin{figure}[t!]
\centering
\includegraphics[width=5.6cm]{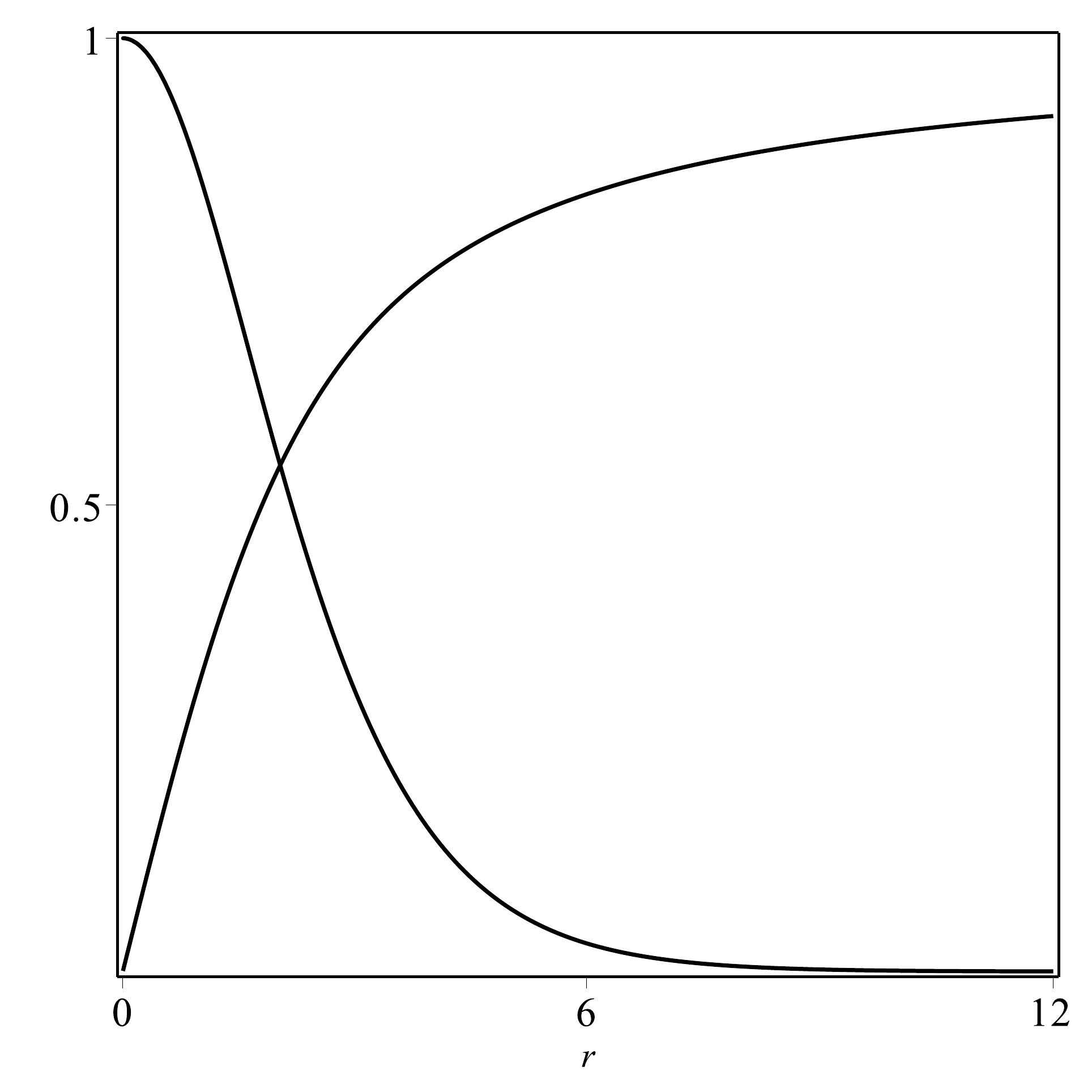}
\caption{The solutions in Eq.~\eqref{solstd}. Here, $K(r)$ is represented by the descending line and $H(r)$ by the ascending one.}
\label{fig1}
\end{figure}
%%%%%%%%%%%%%%%%%%%%%%%

In order to illustrate the above results, from now on we work with dimensionless fields, keeping in mind that the rescaling
\be
\begin{aligned}
	\phi^a &\to \eta \phi^a, & A^a_\mu &\to \eta A^a_\mu, & r &\to (g\eta)^{-1} r, & \LL &\to g^2\eta^4 \LL
\end{aligned}
\ee 
can be done straightforwardly. We then set $g,\eta=1$ and consider only the upper signs in the first order equations \eqref{fom}, which lead to $H(r)\geq0$, for simplicity.

Before going on, let us firstly review the standard model, which was proposed in Refs.~\cite{thooft,polyakov}. It is recovered with
$P(|\phi|)=1$. In this case, the first order equations \eqref{fom} become
\bes\label{foms}
\bal
H^\prime &= \frac{(1-K^2)}{r^2},\\
K^\prime &= -HK.
\eal
\ees
Their solutions are well known \cite{ps}; they are given by
\be\label{solstd}
H(r)= \coth(r) - \frac{1}{r} \quad\text{and}\quad K(r) = r\,\textrm{csch}(r),
\ee
and are displayed in Fig.~\ref{fig1}.

%%%%%%%%%%%%%%%%%%%%%%%
\begin{figure}[t!]
\centering
\includegraphics[width=5.6cm]{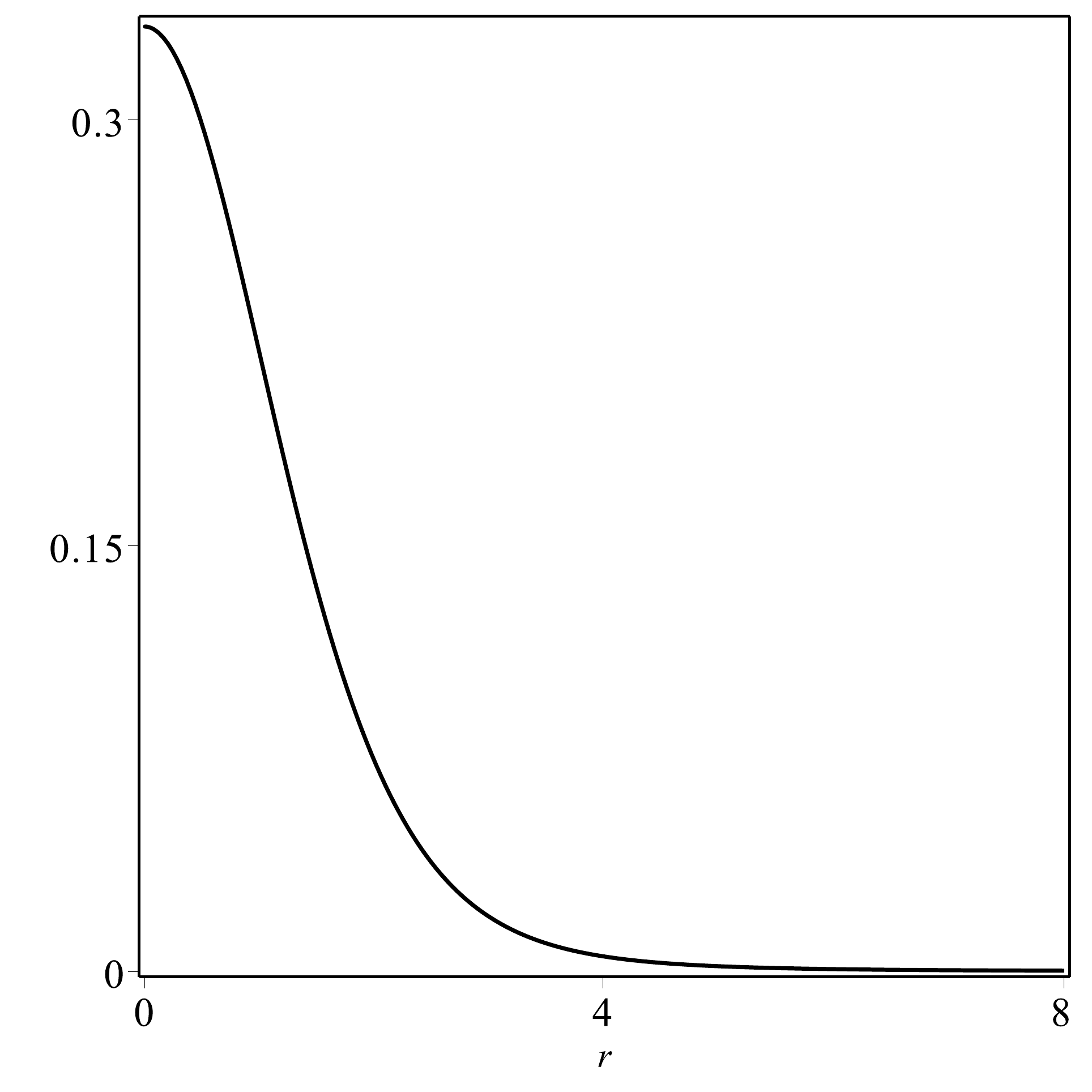}
\includegraphics[width=5cm]{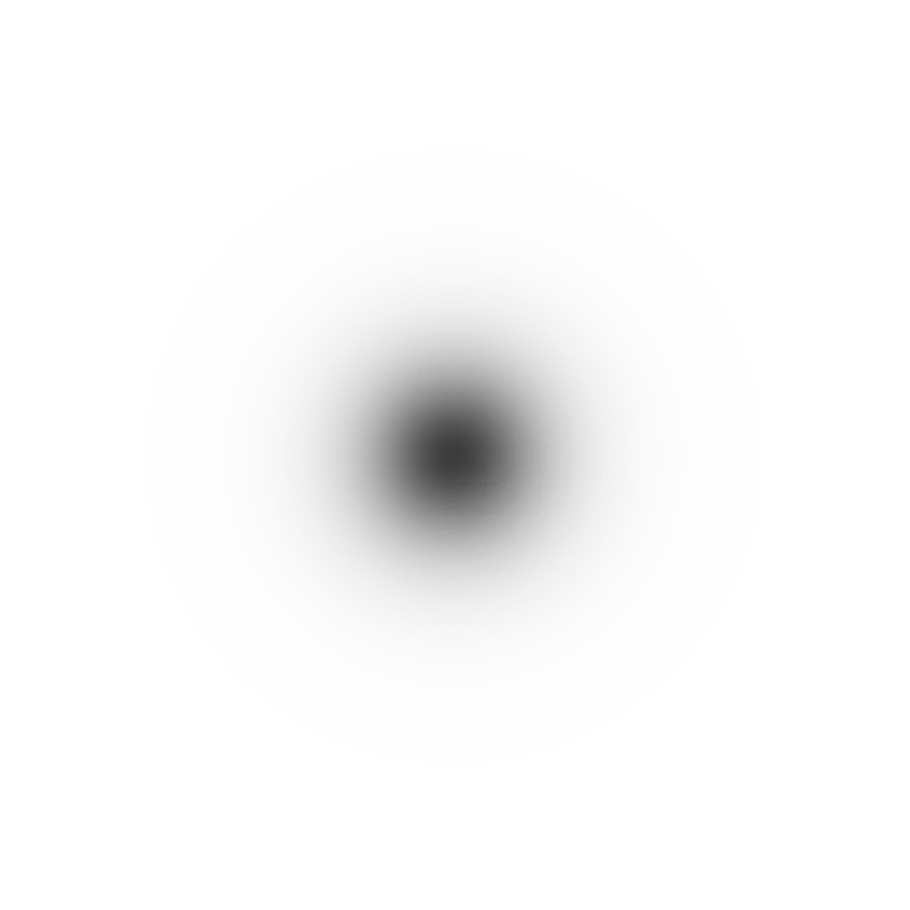}
\caption{In the top panel, we display the energy density \eqref{rhostd}. In the bottom panel, a planar section of the energy density passing through the center of the structure is shown, with the darkness being related to the intensity of the energy density.}
\label{fig2}
\end{figure} 
%%%%%%%%%%%%%%%%%%%%%%%

To calculate the energy density, we consider $P(|\phi|)=1$ in Eq.~\eqref{rhom}, which becomes
\be
\begin{split}
	\rho &= \frac{2{K^\prime}^2}{r^2} + {H^\prime}^2\\
	     &= \frac{2H^2K^2}{r^2} + \frac{(1-K^2)^2}{r^4}.
\end{split}
\ee
By substituting the solutions \eqref{solstd} in the above equation, we get
\be\label{rhostd}
\rho(r) = \frac{\left(r^2\,\textrm{csch}^2(r)-1\right)^2}{r^4} + \frac{2\,\textrm{csch}^2(r)\left(r\coth(r)-1\right)^2}{r^2}.
\ee
In Fig.~\ref{fig2} we depict the above energy density. We also display, in the same figure, the planar section of the energy density passing through the center of the structure for $r\in[0,8]$. This is the standard monopole, which is here revisited for later comparison with the novel structures to be investigated below.

%%%%%%%%%%%%%%%%%%%%%%%%%%%%%%%%%%%%%
\section{New models}\label{sec3}

\begin{figure}[t!]
\centering
\includegraphics[width=6cm]{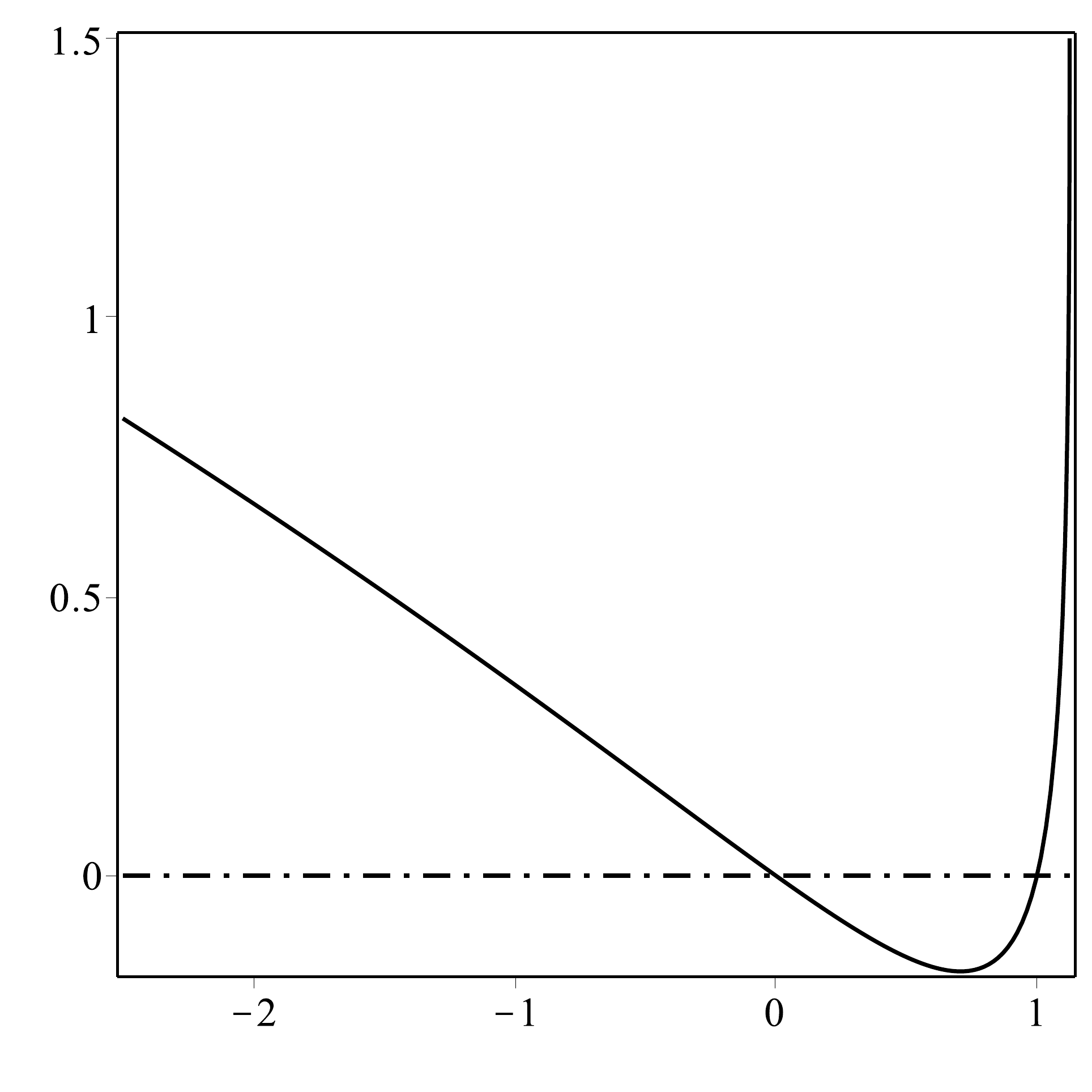}
\caption{The behavior of $\beta$ as a function of $\alpha$.}
\label{fig3}
\end{figure}

Let us now search for other models in the context of the Lagrangian density \eqref{lmodel}, with the motivation to find solutions that engender new features. We guide ourselves by the standard model and consider $P(|\phi|)=|\phi|^\alpha,$ with $\alpha$ being a real parameter that leads us back to the standard case for $\alpha=0$. The first order equations \eqref{fom} become
\bes\label{foms}
\bal
H^\prime &= \frac{H^\alpha (1-K^2)}{r^2},\\
K^\prime &= -H^{{(1-\alpha})}K.
\eal
\ees
We study the behavior of $H$ and $K$ near the origin. They admit the expansion $H(r)\propto r^s$ and $1-K(r)\propto r^{(1+s-s\alpha)}$, and compatibility with the first order equations leads to the values $s=2$ when $\alpha=1$, and 
\be
s=\frac{-1+\sqrt{9-8\alpha}}{2(1-\alpha)},
\ee
for $\alpha\neq 1$. To keep $s$ real we have to impose that $\alpha\leq9/8$. In both cases, the energy density \eqref{rhom} near the origin
behaves as $\rho(r)\propto r^\beta$, where
\be
\beta=\frac{(2-\alpha)\sqrt{9-8\alpha}+5\alpha-6}{2(1-\alpha)}.
\ee
We display $\beta=\beta(\alpha)$ in Fig.~\ref{fig3}, to show that for $\alpha=0$ (the standard case) and for $\alpha=1$, the energy density has a regular behavior at the origin, being distinct positive constants. In the region $\alpha \in (0,1)$, it has a singular behavior at the origin, and for $\alpha< 0$ and $\alpha\in (1,9/8]$ it vanishes as $r$ goes to zero. 

We have studied several distinct possibilities, and noted that for $\alpha<0$, the monopole starts gaining an internal structure, a hole that enlarges as $\alpha$ decreases toward lower and lower (negative) values. A similar situation occurs for $\alpha\in (1,9/8]$. It is a small monopole at $\alpha=1$, and it starts gaining a very small hole at its core, which enlarges very slowly as $\alpha$ increases toward the value $9/8$. These results suggest that we investigate the case $\alpha=1$ explicitly, which is similar to the standard case, and another one, which we choose  to be $\alpha=-2$, to illustrate the possibility of obtaining a monopole with the energy density vanishing at its core. We also consider the case
$\alpha=9/8$, in order to further illustrate our findings.

%%%%%%%%%%%%%%%%%%%%%%%
%%%%%%%%%%%%%%%%%%%%%%%

\subsection{Small Monopole}
As our first example, we consider $\alpha=1$, that is,
\be\label{p1}
P(|\phi|) = |\phi|.
\ee
In this case, the first order equations \eqref{fom} take the form
\bes\label{fo1}
\bal
H^\prime &= \frac{H\,(1-K^2)}{r^2},\\
K^\prime &=- K.
\eal
\ees

%%%%%%%%%%%%%%%%%%%%%%%
\begin{figure}[t!]
\centering
\includegraphics[width=5.6cm]{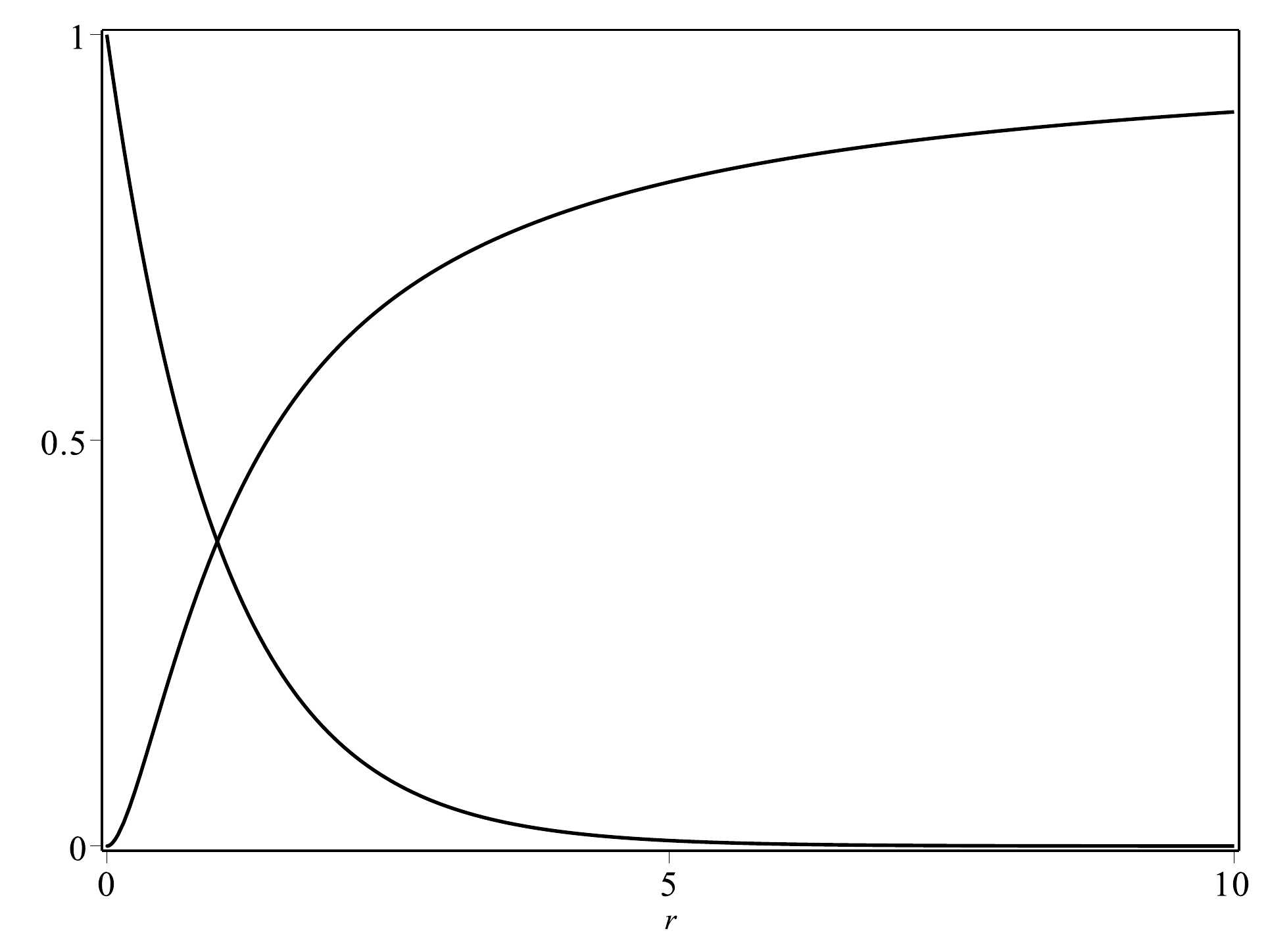}
\caption{The solutions of the first order equations \eqref{fo1}. Here, $K(r)$ is represented by the descending line and $H(r)$ by the ascending one.}
\label{fig4}
\end{figure} 
%%%%%%%%%%%%%%%%%%%%%%%

We have been able to find analytical solutions for the above equations. They are given by
\be\label{sol1}
H(r) = e^{-\frac{e^{2r}-1}{r\,e^{2r}} - 2\textrm{Ei}(1,2r)}, \quad\quad K(r) = e^{-r},
\ee
where $\rm{Ei}(a,z)$ denotes the Exponential Integral function. We highlight here that it is not a simple task to find analytical solutions in generalized models. Even though some papers have dealt with this issue before (see, e.g., Refs.~\cite{mono1,mono2,mono3}), they have provided the magnetic permeability in terms of the coordinate $r$; the starting model, which should depend on the field is unknown. Here, however, we presented the starting model, which is simply driven by the magnetic permeability in Eq.~\eqref{p1}, and have obtained the explicit solutions, which are depicted in Fig.~\ref{fig4}.

%%%%%%%%%%%%%%%%%%%%%%%
\begin{figure}[t!]
\centering
\includegraphics[width=5.6cm]{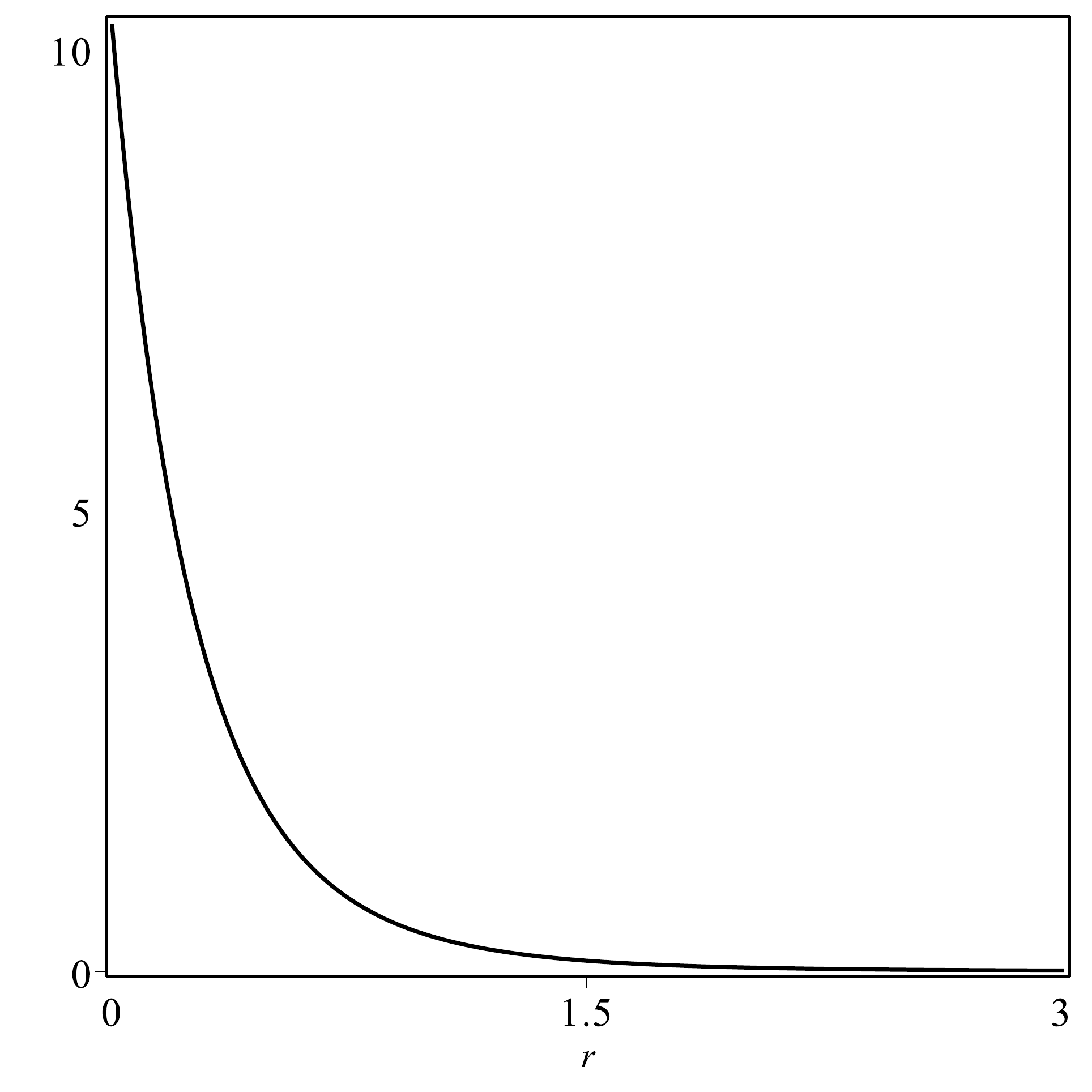}
\includegraphics[width=5cm]{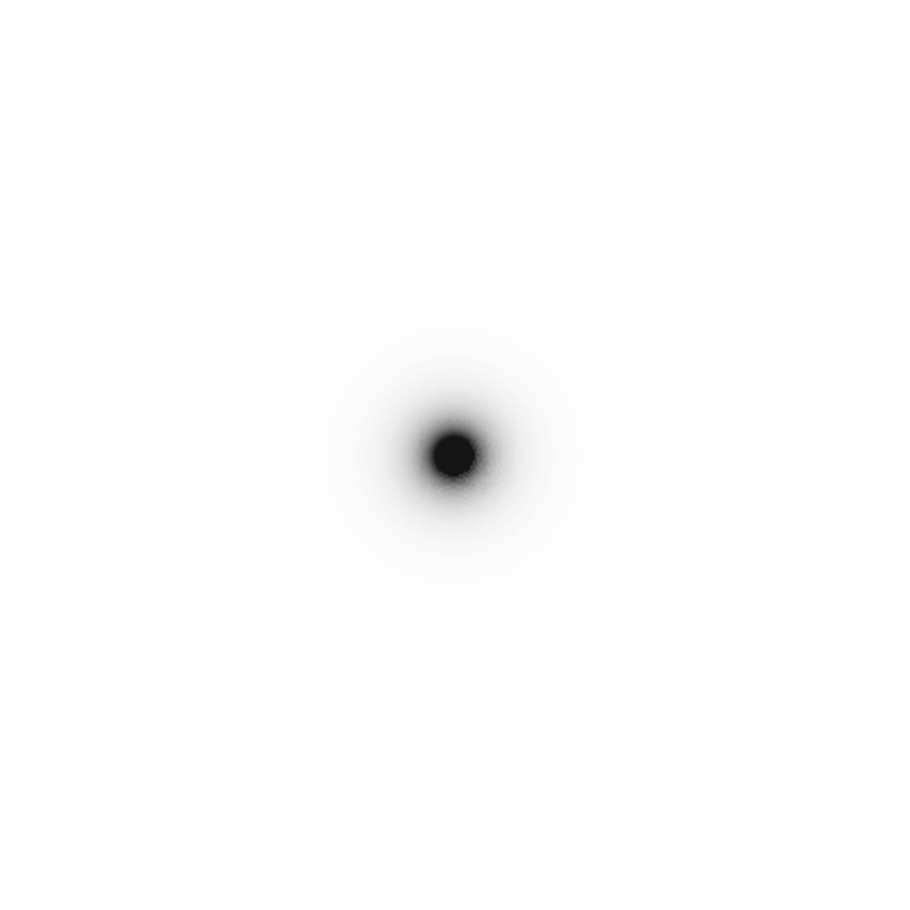}
\caption{In the top panel, we display the energy density \eqref{rho1}. In the bottom panel, a planar section of the energy density passing through the center of the structure is shown, with the darkness being qualitatively related to the intensity of the energy density.}
\label{fig5}
\end{figure} 
%%%%%%%%%%%%%%%%%%%%%%

In order to calculate the energy density, we combine Eqs.~\eqref{p1} and \eqref{rhom} to get
\be
\begin{split}
\rho &= \frac{2H{K^\prime}^2}{r^2} + \frac{{H^\prime}^2}{H}\\
&= \frac{2HK^2}{r^2} + \frac{H\left(1-K^2\right)^2}{r^4}.
\end{split}
\ee
By combining this with the analytical solutions in Eq.~\eqref{sol1}, we get
\be\label{rho1}
\rho=\frac{1}{r^4}\rho_1(r)\,  e^{-\frac{1}{r}\rho_2(r)},
\ee
where
\be 
\rho_1(r)={\left(e^{2r}-1\right)^2 + 2r^2e^{2r}},
\ee
and
\be  
\rho_2(r)={1 -e^{2r}+ 2r\,\textrm{Ei}(1,2r) + 4r^2}.
\ee

In Fig.~\ref{fig5} we display the above energy density; it presents a cusp around the origin and vanishes rapidly. This motivated us to display a planar section of the energy density passing through the center of the structure for $r\in[0,8]$; that is, we keep the same scale used to depict the bottom panel in Fig.~\ref{fig2}, to ease comparison with the size of the standard monopole. We can see from the two energy densities, illustrated in Figs.~\ref{fig2} and \ref{fig5}, that the new structure is smaller and denser. Due to this fact, we call the novel structure the small monopole. One can integrate the energy density to show that the energy is $E=4\pi$, in accordance with Eq.~\eqref{ebogo} for $\eta,g=1$.

\subsection{Hollow Monopole}
\begin{figure}[t!]
\centering
\includegraphics[width=5.6cm]{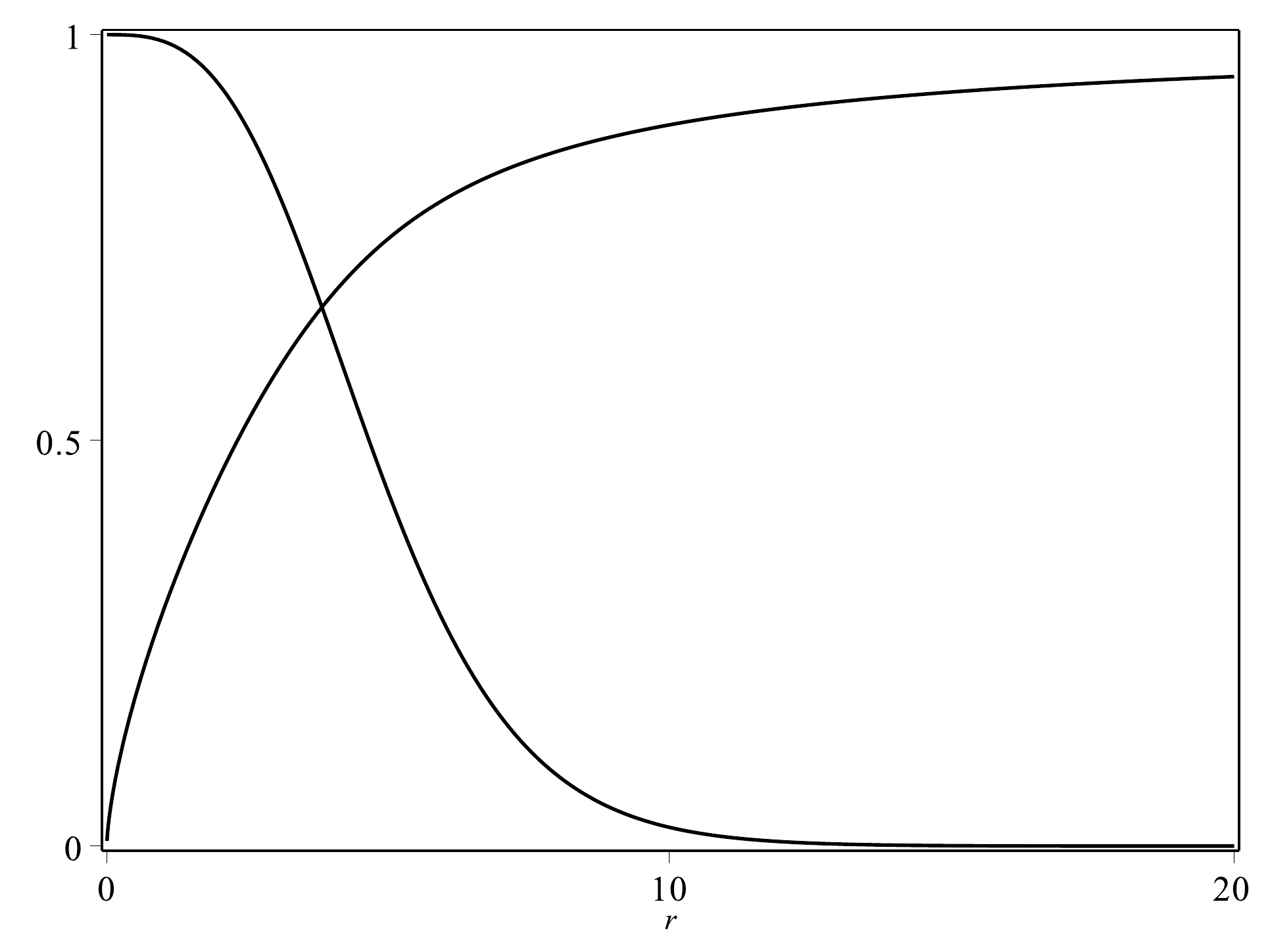}
\caption{The solutions of the first order equations \eqref{fo2}. Here, $K(r)$ is represented by the descending line and $H(r)$ by the ascending one.}
\label{fig6}
\end{figure}

\begin{figure}[t!]
\centering
\includegraphics[width=5.6cm]{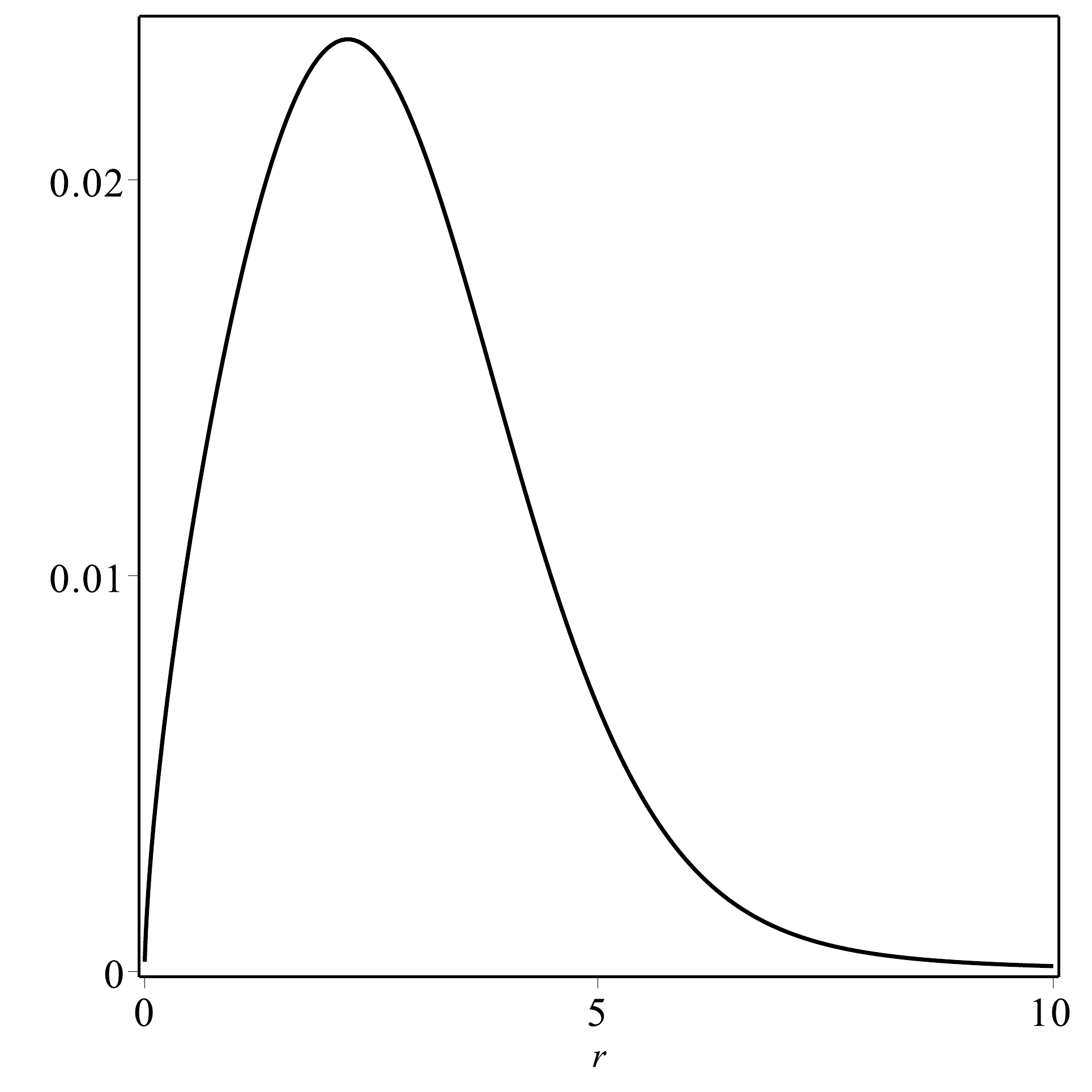}
\includegraphics[width=5cm]{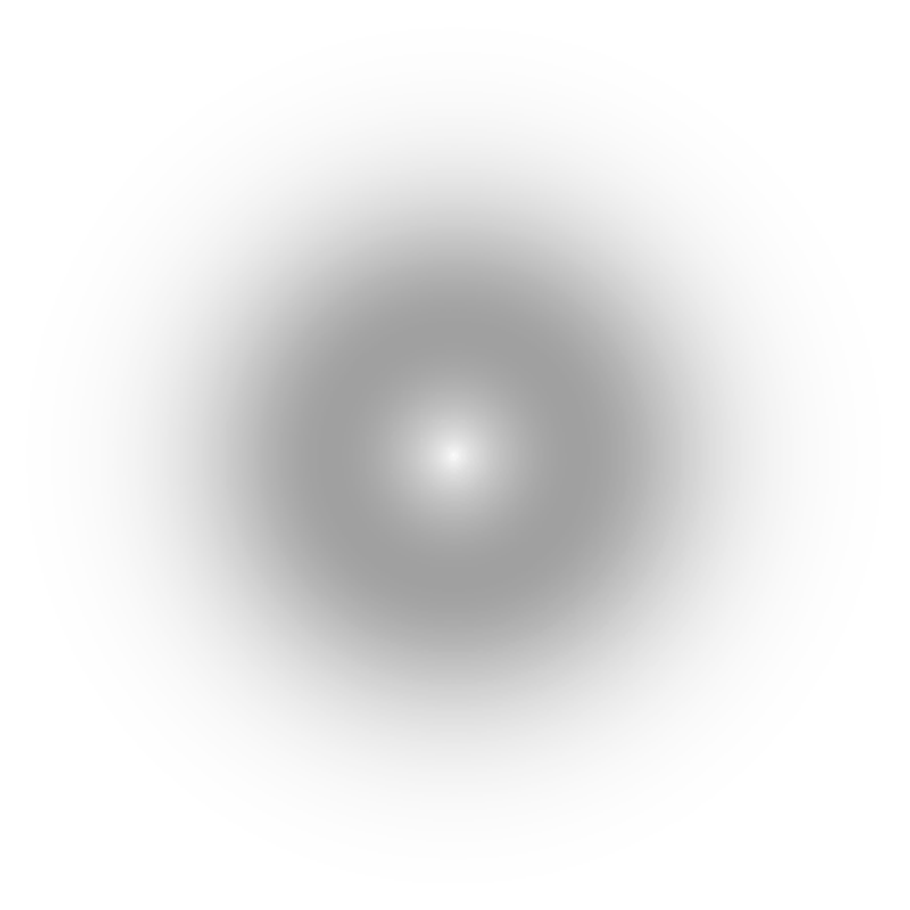}
\caption{In the top panel, we display the energy density \eqref{rho2} for the solutions of the first order equations \eqref{fo2}. In the bottom panel, a planar section of the energy density passing through the center of the structure is shown, with the darkness being qualitatively related to the intensity of the energy density.}
\label{fig7}
\end{figure}

The second example arises with $\alpha=-2$, with the magnetic permeability controlled by the function
\be\label{p2}
P(|\phi|) = \frac{1}{|\phi|^2}.
\ee
In this case, the first order equations are given by
\bes\label{fo2}
\bal
H^\prime &=\frac{(1-K^2)}{r^2H^2},\\
K^\prime &=- H^3K.
\eal
\ees
Differently from the previous model, here we could not find analytical solutions. However, we can estimate their behavior near the origin by taking $H(r) = H_o(r)$ and $K(r) = 1 - K_o(r)$ up to the lowest order, described by $H_o$ and $K_o$. This leads to $H_0(r) \propto r^{2/3}$ and $K_0(r) \propto r^3$. A similar approach can be used to evaluate their asymptotic behavior, considering $H(r)=1-H_{asy}$ and $K(r) = K_{asy}$. By doing so, we get $H_{asy}\propto 1/r$ and $K_{asy}\propto e^{-r}$. Making use of numerical methods, we solve the first order equations \eqref{fo2} and depict the profile of the functions $H(r)$ and $K(r)$ in Fig.~\ref{fig6}. We see that, differently from the previous model, the function
$K(r)$ is approximately constant near the origin and goes to its asymptotic value faster than $H(r)$.
%%%%%%%%%%%%%%%%%%%%%%%
%%%%%%%%%%%%%%%%%%%%%%%
In order to calculate the energy density, we make use of Eq.~\eqref{rhom} to write
\be\label{rho2}
\begin{split}
	\rho_m &= \frac{2P{K^\prime}^2}{r^2H^2} + H^2{H^\prime}^2\\
	       &= \frac{2H^4K^2}{r^2} + \frac{(1-K^2)^2}{r^4H^2}
\end{split}
\ee
We substitute the numerical solutions of Eqs.~\eqref{fo2} in the above equation and display the energy density in Fig.~\ref{fig7}. This model presents an interesting feature: its energy density vanishes in the center of the monopole. In this sense, due to the absence of matter at the origin, we say that the monopole presents an internal structure. This characteristic motivated us to plot a planar section passing through the center of the structure for $r\in[0,8]$, also in Fig.~\ref{fig7}. Note that the scale used to depict the planar section in Fig.~\ref{fig7} is the same as in Figs.~\ref{fig2} and \ref{fig5}, to ease comparison. To show that the last structure is larger then the standard monopole. This figure also allows us to conclude that the monopole has a hole at its center, so we called it the hollow monopole, with the hole and the energy density contributing to make the structure larger than the previous ones. Notice that the new features were obtained here in a purely $SU(2)$ model, without the presence of an additional symmetry as in Ref.~\cite{monopoleint}.

\begin{figure}[t!]
\centering
\includegraphics[width=4.2cm]{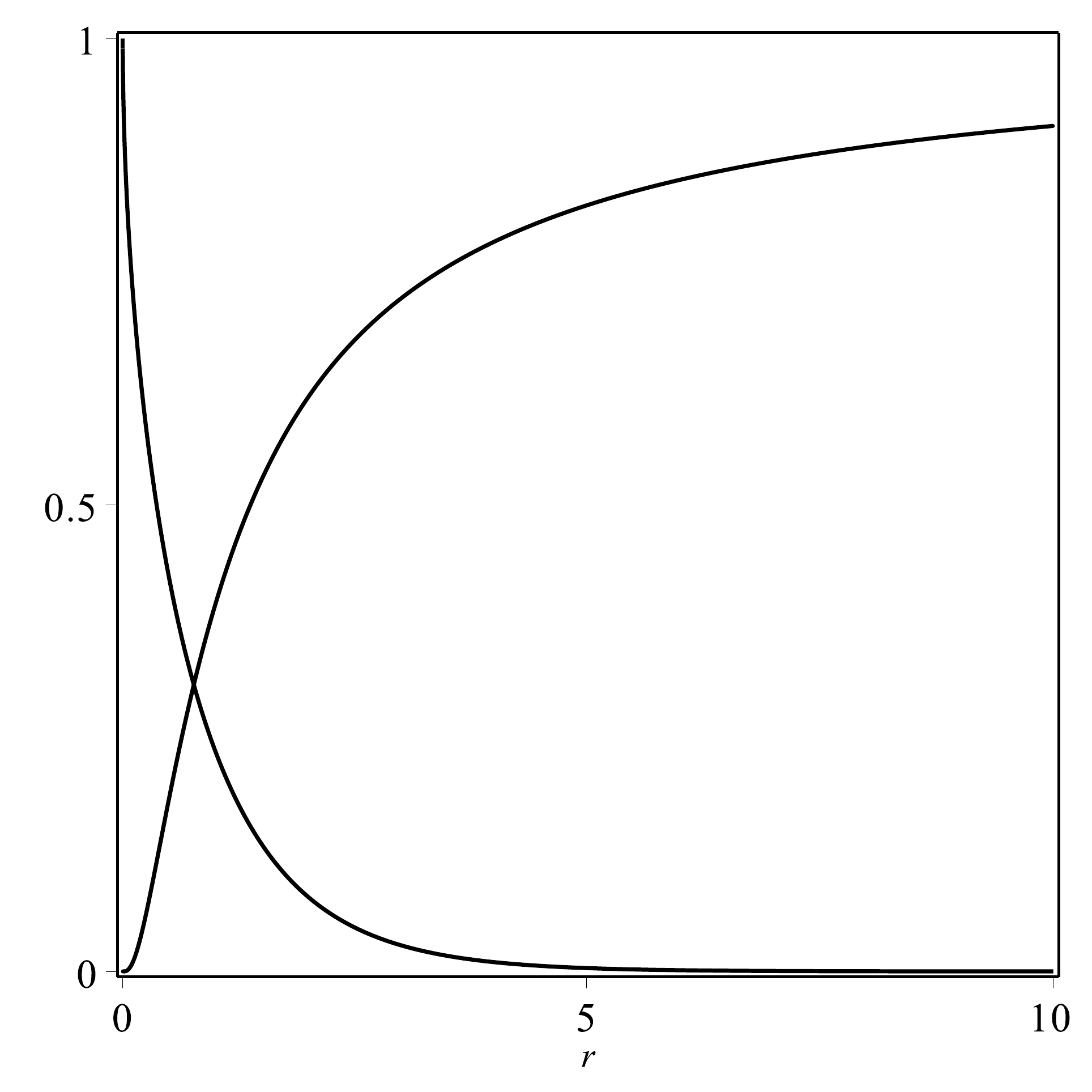}
\includegraphics[width=4.2cm]{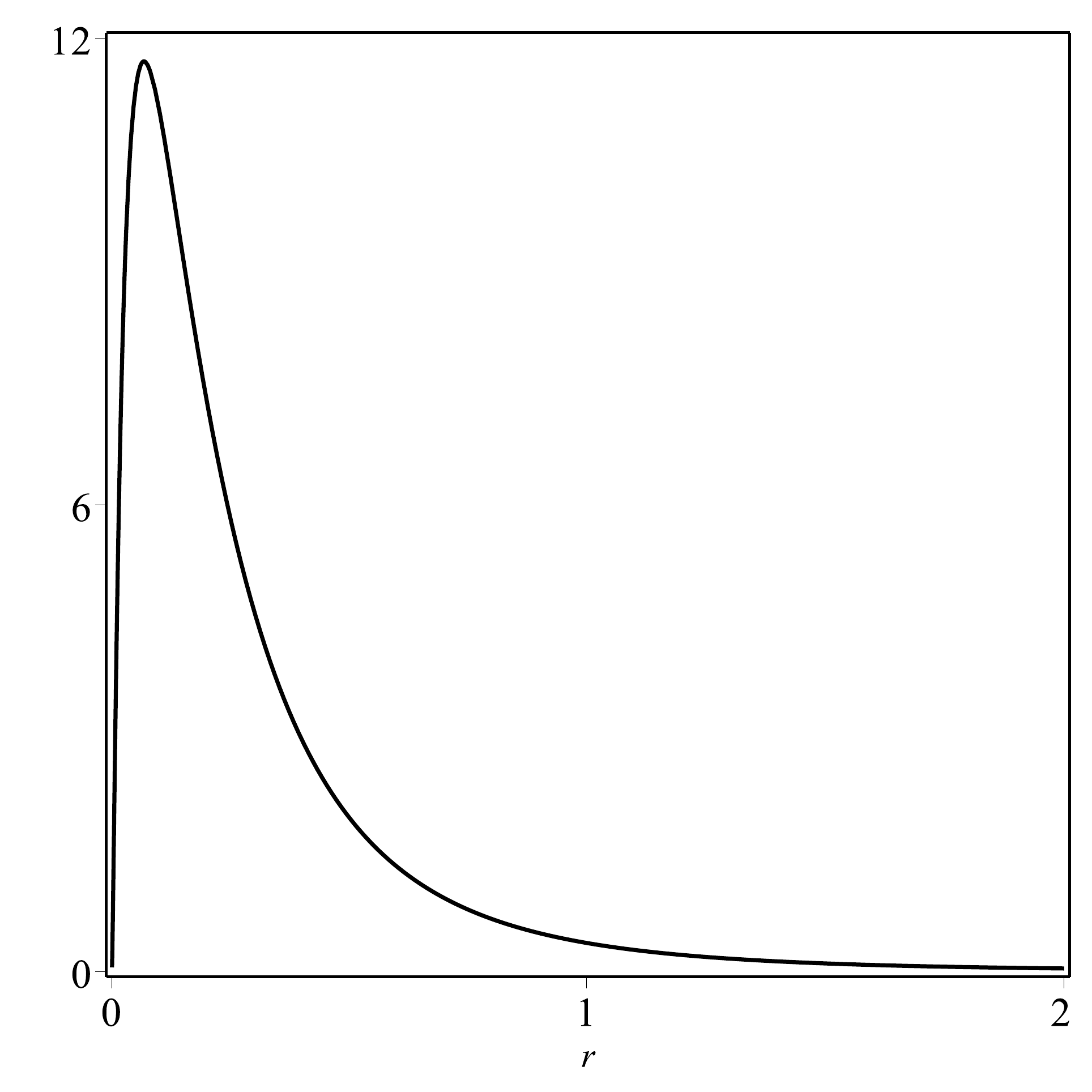}
\includegraphics[width=5cm]{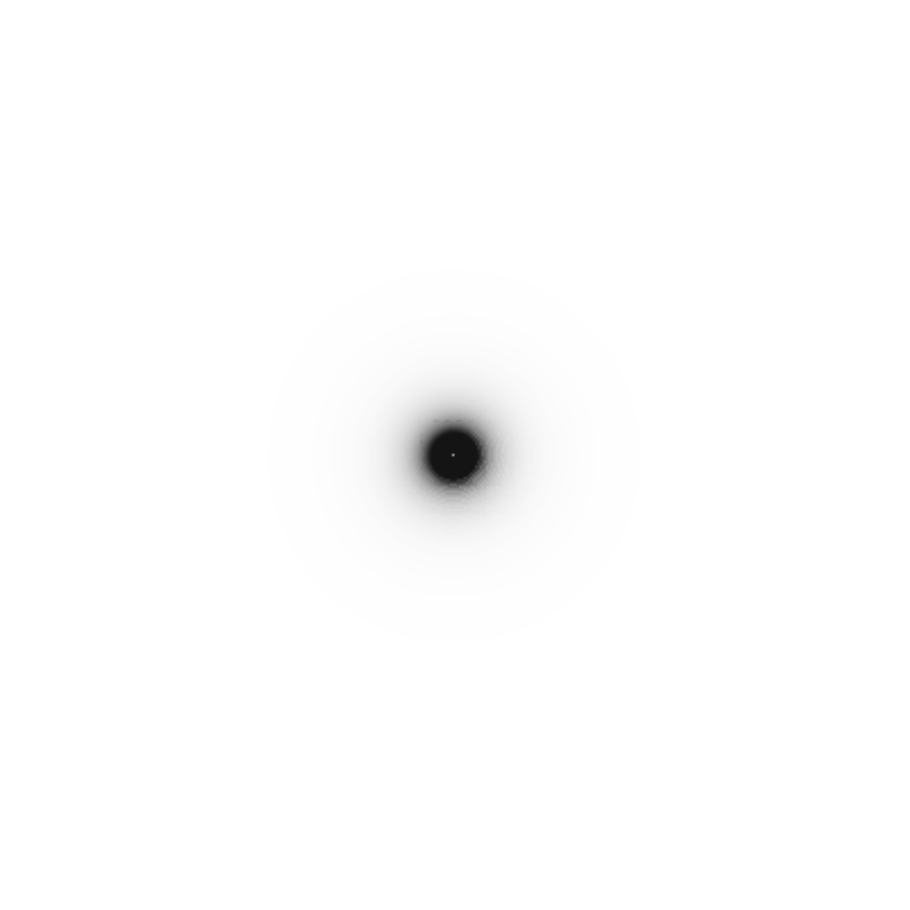}
\caption{In the top panel, we display the solutions (left) and the energy density (right) for the model with $\alpha=9/8$. In the bottom panel, a planar section of the energy density passing through the center of the structure is shown, with the darkness being qualitatively related to the intensity of the energy density.}
\label{fig8}
\end{figure}

The case of $\alpha=9/8$ was also considered. The investigation follows as in the previous cases, and the results are displayed in Fig.~\ref{fig8}. There, in the bottom panel we also use $r\in[0,8]$, as we did before in Figs.~\ref{fig2}, \ref{fig5} and \ref{fig7}. We see that the monopole is still small, and it gains an almost invisible hole at its center.

%%%%%%%%%%%%%%%%%%%%%%%
%%%%%%%%%%%%%%%%%%%%%%%

%%%%%%%%%%%%%%%%%%%%%%%
%%%%%%%%%%%%%%%%%%%%%%%

%%%%%%%%%%%%%%%%%%%%%%%%%%%%%%%%%
\section{Conclusions}\label{sec4}

In this work, we have investigated the $SU(2)$ model described by the Lagrangian density \eqref{lmodel}, with the focus on the construction of magnetic monopoles. We have presented the general properties of the model, including the equations the fields should obey and the energy density. Despite the presence of spherical symmetry, the equations of motion are coupled second order differential equations, hard to be solved. For this reason, we have used the BPS procedure to minimize the energy and obtain first order differential equations that solve the equations of motion. 

We illustrated the general features of the model by properly choosing the magnetic permeability. Some distinct possibilities were considered, one with $\alpha=1$, that gives rise to analytic solutions, with the corresponding energy density shrinking around its core and concentrating significantly, when compared with the standard structure obtained in Refs.~\cite{thooft,polyakov} and reviewed here for comparison. We called the novel solution the small monopole, because it is really smaller than the standard monopole, as it appears very clearly when one compares the two Figs.~\ref{fig2} and \ref{fig5}. The second model was defined with $\alpha=-2$; it behaves differently, and we have been unable to find analytical solutions. However, the corresponding energy density was shown to present a hole in its core, a feature that motivated us to call it the hollow monopole. The novel magnetic structure, the hollow monopole engenders energy density which spreads out from the hole and makes it at least twice as large as the standard monopole, as one sees from Figs.~\ref{fig2} and \ref{fig7}. We have also investigated the case of $\alpha=9/8$, to find the small monopole with an internal structure, with a hole at its core. This situation is also of interest, although the effect of the presence of the hole is very subtle and hard to be identified.   

We remark here that both the small and the hollow monopole were obtained for the class of models controlled by the $SU(2)$ model, differently from the case studied before in Ref.~\cite{monopoleint}, where the $SU(2)$ symmetry is enlarged to $SU(2)\times Z_2$, with the additional $Z_2$ symmetry used to accommodate a neutral scalar field that drives the magnetic permeability to allow for the presence of internal structure. As we have seen, the two possibilities are quite different, but both admit the attainment of first order differential equations that solve the corresponding equations of motion and the consequent construction of monopole solutions with internal structure. However, we learn from the results of the current work that the extension of the $SU(2)$ symmetry to include the neutral degree of freedom is not mandatory for the presence of internal structure.  

We are now investigating the inclusion of fermions to explore their behavior in the background of such novel structures, a possibility that can be implemented under the lines of Refs.~\cite{BM1,BM2}. The inclusion of fermions may also allow for supersymmetric extensions, to see how the supersymmetry works to reproduce the first order equations presented in this work. Another interesting issue concerns the study of an enlarged model with the $SU(2)\times SU(2)$ symmetry, with the purpose of describing how the monopoles interact with one another, advancing toward the process of dimerization of monopoles \cite{MD}. The interest in the $SU(2)\times SU(2)$ symmetry is also connected with the possibility of describing the system in the context of one (visible) sector interacting with the other (hidden) sector. This can be implemented along the lines of the studies described before in \cite{hidden3,hidden4,hidden5,BLMM}, and in references therein. We hope to report on these issues in the near future.

\acknowledgements{The work is partially supported by CNPq (Brazil), by the projects FIS2014-57387-C3-1-P, FIS2017-84440-C2-1-P (AEI/FEDER, EU) and  SEJI/2017/042 (Generalitat Valenciana), the Consolider Program CPANPHY-1205388, and the Severo Ochoa grant SEV-2014-0398 (Spain). DB acknowledges support from project CNPq:306614/2014-6 and MAM acknowledges support from project CNPq:140735/2015-1. GJO is supported by the Ramon y Cajal project RYC-2013-13019.}
%%%%%%%%%%%%%%%%%%%%%%%%%%%%%%%%%

\end{document}